\documentstyle[12pt,a4,epsf]{article}

\def\tr{\mathop{\rm tr}\nolimits}
\def\Tr{\mathop{\rm Tr}\nolimits}

\makeatletter
\@addtoreset{equation}{section}
\makeatother

\newcommand{\VEV}[1]{\left\langle #1 \right\rangle}

\newcommand{\Mp}{M_P}

\begin{document}
\begin{titlepage}

\begin{flushright}
hep-ph/0104200\\
KUNS-1715\\
\today
\end{flushright}

\vspace{4ex}

\begin{center}
{\large \bf
Neutrino Masses, Anomalous $U(1)$ Gauge Symmetry \\
and Doublet-Triplet Splitting
}

\vspace{6ex}

\renewcommand{\thefootnote}{\alph{footnote}}

Nobuhiro Maekawa\footnote
{e-mail: maekawa@gauge.scphys.kyoto-u.ac.jp
}

\vspace{4ex}
{\it Department of Physics, Kyoto University,\\
     Kyoto 606-8502, Japan}\\
\end{center}

\renewcommand{\thefootnote}{\arabic{footnote}}
\setcounter{footnote}{0}
\vspace{6ex}

%--------------------<<   abstract   >>--------------------
\begin{abstract}

We propose an attractive scenario of grand unified theories
in which doublet-triplet splitting is naturally realized in
$SO(10)$ unification using the Dimopoulos-Wilczek mechanism. 
The anomalous $U(1)_A$ gauge symmetry plays an essential role
in the doublet-triplet splitting mechanism. It is interesting
that the anomalous $U(1)_A$ charges determine the unification 
scale and mass spectrum of additional particles, as well as the 
order of the Yukawa couplings of quarks and leptons. For the neutrino 
sector, bi-maximal mixing angles are naturally obtained, and 
proton decay via dimension 5 operators is suppressed.
It is suggestive that the anomalous $U(1)_A$ gauge symmetry 
motivated by superstring theory effectively solves the two biggest 
problems in grand unified theories, the fermion mass hierarchy problem 
and doublet-triplet splitting problem.

\end{abstract}

\end{titlepage}

%--------------------<<   section    >>--------------------

\section{Introduction}

The Standard Model is consistent with all present experiments.
However, there are many reasons for thinking that it is not the final 
theory;
for example, it does not explain the anomaly cancellation 
between quarks and leptons, the hierarchies of gauge and Yukawa couplings, 
charge quantization, etc.
Therefore we have a strong motivation for examining the idea of grand 
unified theories (GUT)
\cite{georgi}, in which the quarks and leptons are beautifully unified
in several multiplets in a simple gauge group. Three gauge groups in the
Standard Model are unified into a simple gauge group at a GUT scale that 
is considered to be just below the Planck scale. Once we accept a higher
scale than the weak scale, it is one of the most promising ways to 
introduce 
supersymmetry (SUSY) around the weak scale to stabilize the weak scale.
We are thus led to examine SUSY GUT
\cite{SUSYGUT}.

However, it is not easy to obtain a realistic SUSY GUT.
One of the reasons is that it is difficult to obtain a realistic 
fermion mass 
pattern in a simple way, because a unified multiplet introduces strong 
constraints 
on the Yukawa couplings of quarks and leptons.
Moreover, one of the most difficult obstacles in building a realistic GUT is 
the ``doublet-triplet (DT) splitting problem". Generally, a fine-tuning is 
required to 
obtain the light $SU(2)_L$ doublet Higgs multiplet of the weak scale while 
keeping 
the triplet Higgs sufficiently heavy to suppress the dangerous proton decay. 

For the former problem, by using the information on neutrino masses
obtained in recent neutrino experiments
\cite{SK}, there are several
impressive papers
\cite{Sato,Nomura,Bando,Barr,Shafi}
attempting to explain the order of the Yukawa couplings,
though most of these treatments need tuning parameters to explain the large 
mixing angle for the atmospheric neutrino. It is natural to examine 
$SO(10)$ and higher gauge groups because all quarks and leptons, 
including the
right-handed neutrino, can be unified into a single multiplet. This is 
important to investigate neutrino masses.

There have been several attempts to avoid the latter problem
\cite{DTsplitting,DW}.
One of the most promising ways to realize DT splitting in the $SO(10)$ 
SUSY GUT 
is using the Dimopoulos-Wilczek (DW) mechanism
\cite{DW,BarrRaby,Chako,complicate}. If the adjoint field $A$ of 
$SO(10)$ 
has a vacuum expectation value (VEV) $\VEV{A}=i\tau_2\times {\rm diag}
(v,v,v,0,0)$, then 
$SO(10)$
is broken to $SU(3)_C\times SU(2)_L\times SU(2)_R\times U(1)_{B-L}$,
and the 
VEV can impart masses on the triplet Higgs but not to the doublet Higgs.
Unfortunately, in order to realize
the DW mechanism, a rather complicated Higgs structure is required
\cite{complicate}. The reason
is simple: The DW mechanism works essentially in a larger rank
unified gauge group
(like $SO(10)$ GUT) than that of the standard gauge group. For example, 
since the adjoint field in $SU(5)$ GUT is traceless and the rank is 
the same
as that of the standard gauge group, 
it is impossible to realize the DW mechanism. On the other hand, we have to 
introduce VEVs of spinors $C$ and $\bar C$ or other multiplets to 
break the 
remaining gauge 
group to the standard gauge group, because the adjoint VEV does not 
reduce the 
rank of the $SO(10)$ gauge group. If the VEV of the spinor appears in the 
equation 
of motion that determines the adjoint VEV, the VEV of the adjoint 
field generally deviates from the form required for the DW mechanism. 
On the other hand,
if the adjoint and spinor Higgs sectors are not coupled to each other 
in the
superpotential, then pseudo Nambu-Goldstone (PNG) fields 
$({\bf 3},{\bf 2})_{1/6}+({\bf 3},{\bf 1})_{-2/3}$+h.c. of 
$SU(3)_C\times SU(2)_L\times U(1)_Y$
appear. 

To avoid this problem, the adjoint field must couple to the spinor
to obtain the mass of the PNG fields, retaining the DW mechanism.
It is not obvious that this is possible, but in fact it is.
%Is it possible? The answer is Yes. 
This is possible because, for the equation of 
motion,
the first derivative of the superpotential is important, while for the 
mass term, the second derivative is essential. However, usually realizing
this situation 
requires a rather complicated Higgs sector. Recently Chacko and Mohapatra 
proposed a simpler model
\cite{Chako}, in which they introduce two ${\bf 45}$,
one ${\bf 54}$, a pair consisting of ${\bf 16}$ and ${\bf \overline{16}}$, 
and two ${\bf 10}$
just for the Higgs sector. 
Several years ago, Barr and Raby
\cite{BarrRaby} examined a minimal DT splitting model
that includes a single ${\bf 45}$, two pairs of ${\bf {16}}$ and 
${\bf \overline{16}}$ 
and two ${\bf 10}$ for the Higgs sector. This simple model is very 
attractive.
However, it requires the introduction of many singlets whose VEVs are not 
determined
classically and  must be given by hand.
Moreover, in their model, dangerous terms are not forbidden by symmetry.
Once the mass term $A^2$ and the non-renormalizable term $A^4$, 
which are essential for 
their model, are allowed, there is no reason to forbid 
higher power terms $A^{2n}$. With
these terms, because many (infinitely many)
degenerate undesired vacua appear, 
it is unnatural to obtain the desired DW vacuum.

In this paper, we propose a more attractive DT splitting scenario 
in which the GUT scale
is automatically determined and the higher terms are naturally forbidden.
In this scenario, the anomalous $U(1)_A$ gauge symmetry plays an essential 
role.
Using this mechanism, a GUT with realistic Yukawa couplings can be 
constructed
in a simple way. This model has interesting quark and lepton
 mass matrix structure, which predicts bi-maximal mixing in the neutrino sector.

%--------------------<<   section    >>--------------------
\section{Anomalous $U(1)_A$ gauge symmetry and neutrino masses}
\label{sec:anom}
First let us recall the anomalous $U(1)_A$ gauge symmetry. It is
well known that some low energy effective theories of the string
theory include the anomalous $U(1)_A$ gauge symmetry, which has
non-zero anomalies, such as the pure $U(1)_A^3$ anomaly, mixed
anomalies with other gauge groups $G_a$, and a mixed
gravitational anomaly \cite{U(1)}. These
anomalies are canceled by combining the nonlinear transformation of 
the dilaton chiral supermultiplet $D$ with the gauge transformation 
of the $U(1)_A$ vector supermultiplet $V_A$ as
\begin{eqnarray}
V_A &\to&  V_A
  +\frac{i}{2}\left(\Lambda-\Lambda^\dagger\right),\\
D &\to& D+\frac{i}{2}\,\delta_{GS} \,\Lambda,
\label{trans}
\end{eqnarray}
where $\Lambda$ is a parameter chiral superfield.
This cancellation occurs because the gauge kinetic functions 
for $V_A$ and the other
vector supermultiplets $V_a$ are given by
\begin{equation}
{\cal L}_{\mbox{\scriptsize gauge}}
 = \frac{1}{4} \int d^2\theta \left[\,k_A D \,W_A{}^\alpha
W_A{}_\alpha 
       + k_a D\, W_a{}^\alpha W_a{}_\alpha \,\right] + \mbox{h.c.}, 
\end{equation}
where $W_A{}^\alpha$ and $W_a{}^\alpha$ are the super field strengths of 
$V_A$ and $V_a$, and  
$k_A$ and $k_a$ are Kac-Moody levels of $U(1)_A$ and $G_a$. 
The square of the
gauge coupling is written in terms of the inverse of the VEV of the 
dilaton as $k_a\VEV{D}=1/g_a^2$.

The parameter $\delta_{GS}$ in Eq.~(\ref{trans}) is related to
the conditions for the anomaly cancellations,\footnote
{ 
$C_a\equiv\Tr_{G_a} T(R)\,Q_A$.  Here
$T(R)$ is the Dynkin index of the representation $R$, and we use the
convention in which $T(\mbox{fundamental rep.}) = 1/2$.
}
\begin{equation}
2\pi^2\delta_{GS}\,= \,\frac{C_a}{k_a}
\,= \,\frac{1}{3k_A}\tr {Q_A}^3 
\,= \,\frac{1}{24}\tr Q_A.
\end{equation}
The last equality is required by the cancellation of the mixed
gravitational anomaly. These anomaly cancellations are understood in
the context of the Green-Schwarz mechanism\cite{GS}.

One of the most interesting features of the anomalous $U(1)_A$ gauge
symmetry is that it induces the Fayet-Iliopoulos $D$-term (F-I term)
radiatively\cite{U(1)}. Since the K\"ahler potential $K$ for the
dilaton $D$ must 
be a function of $D+D^\dagger-\delta_{GS}V_A$
for $U(1)_A$ gauge invariance,
the F-I term can be given as 
\begin{equation}
\int d^4\theta \,K(D+D^\dagger-\delta_{GS}V_A)
=  \left(-\frac{\delta_{GS}K'}{2}\right)D_A + \cdots
  \equiv \xi^2 D_A +\cdots,
\label{FIterm}
\end{equation}
where we take the sign of $Q_A$ so that $\xi^2 >0$. 
If the K\"ahler potential for the dilaton is given by
$K=-\ln (D+D^\dagger-\delta_{GS}V_A)$, which can be induced by a stringy
calculation at tree level, $\xi^2$ can be approximated as
\begin{equation}
\xi^2=\frac{g_s^2\tr Q_A}{192\pi^2},
\end{equation}
where $g_s^2=1/\VEV{D}$.
Note that since $\xi^2$ is induced radiatively, the parameter 
$\xi$
is expected to be smaller than the Planck scale. 

When some superfields $\Phi_i$ have anomalous $U(1)_A$ charges
$\phi_i$, the scalar potential becomes
\begin{eqnarray}
V_{\mbox{\scriptsize scalar}}
&= &\frac{g_A^2}{2}
      \left(\sum_i \phi_i |\Phi_i|^2 
            +\xi^2 \right)^2,        
\end{eqnarray}
where $1/g_A^2=k_A \VEV{D}$.
If one superfield has a negative 
anomalous $U(1)_A$ charge, it acquires a non-zero VEV.
Below, we assume the existence of a field $\Phi$ with negative
charge and normalize the anomalous $U(1)_A$ charges so that
$\Phi$ has charge $-1$.    
Then the VEV of the scalar component $\Phi$ is given by
\begin{equation}
\VEV{\Phi}=  \xi \equiv \lambda \Mp, 
\label{VEV_Phi}
\end{equation}
which breaks the anomalous $U(1)_A$ gauge symmetry. 
(Here $\Mp$ is some gravity scale and is usually taken as 
the reduced Planck mass, $1/\sqrt{8\pi G_N}$. In the following,
we use units in which $\Mp=1$.)

Next we discuss the fermion masses.
In general, the Yukawa hierarchy can be explained by introducing a  
flavor dependent $U(1)$ symmetry
\cite{FN,Ibanez,Ramond,Dreiner}.
We can adopt the anomalous $U(1)_A$ gauge symmetry as this $U(1)$
symmetry. Suppose that the Standard Model matter fields $Q_i$,
$U^c_i$, $D^c_i$, $L_i$, $E^c_i$, $H_u$ and $H_d$ have the
anomalous $U(1)_A$ charges $q_i$, $u_i$, $d_i$, $l_i$,
$e_i$, $h_u$ and $h_d$,
%\footnote{
%Through this paper we denote lowercase as the anomalous $U(1)_A$ charge.
%}
\footnote{
Throughout this paper we denote all the superfields with uppercase 
letters
and their anomalous $U(1)_A$ charges with the corresponding lowercase 
letters.
}
which are taken as non-negative integers here.
If the field $\Phi$ with charge $-1$ is a singlet under the
Standard Model gauge symmetry, the superpotential can be written as
\begin{eqnarray}
W \sim
  \Phi^{q_i+u_j+h_u}
    H_uQ_i U^c_j + \cdots. 
\end{eqnarray}
In this paper, for simplicity, we usually do not write $O(1)$ 
coefficients explicitly. 
Since the scalar component of $\Phi$ has the VEV given in 
(\ref{VEV_Phi}),
we obtain the hierarchical mass matrices
\begin{eqnarray}
(M_u)_{ij}
&\sim&
\lambda^{q_i+u_j+h_u} \VEV{H_u}
= V^u_L 
\left(
\begin{array}{ccc}
m_u & & \\
 & m_c & \\
 & & m_t
\end{array}
\right)
V^{u\dagger}_R , 
\label{diag_matrices1}\\
(M_d)_{ij}
&\sim&
\lambda^{q_i+d_j+h_d} \VEV{H_d} 
= V^d_L 
\left(
\begin{array}{ccc}
m_d & & \\
 & m_s & \\
 & & m_b
\end{array}
\right)
V^{d\dagger}_R,
\label{diag_matrices2}
\end{eqnarray}
where 
%$m_f$ are masses of quarks.
the $V^{u,d}_{L,R}$ are $3\times 3$ unitary diagonalizing matrices, and 
$(V^u_L)_{ij} \sim \lambda^{|q_i-q_j|}$,
$(V^u_R)_{ij} \sim \lambda^{|u_i-u_j|}$, and so on. 
The diagonalized masses of quarks, $m_f$, satisfy
 $(m_u)_i\sim \lambda^{q_i+u_i+h_u}\VEV{H_u}$ and
$(m_d)_i\sim \lambda^{q_i+d_i+h_d}\VEV{H_d}$.
The Cabbibo-Kobayashi-Maskawa matrix
\cite{CKM} is given by
\begin{eqnarray}
V_{\rm CKM}
= V^u_L {V^d_L}^{\dagger}
\sim \left(
\begin{array}{ccc}
1&\lambda^{|q_1-q_2|}&\lambda^{|q_1-q_3|}\\
\lambda^{|q_2-q_1|}&1&\lambda^{|q_2-q_3|}\\
\lambda^{|q_3-q_1|}&\lambda^{|q_3-q_2|}&1 
\end{array}
\right) ,
\end{eqnarray}
which is determined solely by the charges of the left-handed quarks,
$q_i$.
The relation $V_{12}V_{23}\sim V_{13}$ can naturally be understood with 
this
mechanism, and if we take $q_i= (3,2,0)$ and $\lambda \sim 0.2$, 
we can obtain the measured value. 

If there are right-handed neutrinos $N_i^c$ with $U(1)_A$ charges $n_i$, 
the Dirac and Majorana 
neutrino masses are also given by 
\begin{eqnarray}
  (M_D)_{ij}&\sim& \lambda^{l_i+n_j+h_u}\VEV{H_u},\\  
  (M_R)_{ij} &\sim& M_m \lambda^{n_i+n_j}.
  \label{M_R}
\end{eqnarray}
Through the see-saw mechanism
\cite{seasaw},
the left-handed neutrino mass matrix is given by
\begin{eqnarray}
  (M_\nu)_{ij} &\sim& \lambda^{l_i+l_j+2h_u}\frac{\VEV{H_u}^2}{M_m}.
  \label{M_N}
\end{eqnarray}
The mixing matrix for the lepton sector \cite{MNS}
is induced as for the quark  sector:
\begin{eqnarray}
V_{\rm MNS}
= V^\nu_L {V^e_L}^{\dagger}
\sim \left(
\begin{array}{ccc}
1&\lambda^{|l_1-l_2|}&\lambda^{|l_1-l_3|}\\
\lambda^{|l_2-l_1|}&1&\lambda^{|l_2-l_3|}\\
\lambda^{|l_3-l_1|}&\lambda^{|l_3-l_2|}&1 
\end{array}
\right).
\end{eqnarray}
This matrix is also determined only by the charges of the left-handed 
leptons, $l_i$.
If we take $l_i=(2,2,2)$, it generally gives large mixing angles among
the three generations and can give the bi-maximal mixing angles.
This is called the `anarchy solution' for large mixing angles in the
neutrino sector
\cite{anarchy}.

Until this point, we have examined only terms with non-negative total 
anomalous
$U(1)_A$ charge, but we also wish to know what happens if the total 
charge becomes negative.
The terms with negative total anomalous $U(1)_A$ charge are forbidden 
by
the anomalous $U(1)_A$ gauge symmetry, while the terms with positive or 
zero charge are allowed, because the negative charge of the singlet 
$\Phi$ can compensate for the positive charge, as discussed above.
The vanishing of the coefficients resulting from the anomalous $U(1)_A$ 
gauge 
symmetry is called  ``SUSY zero" mechanism.
This feature plays an essential role in our mechanism of DT splitting.

In Eq.~(\ref{M_R}), we have to introduce the Majorana mass scale $M_m$
smaller than $\Mp$ by hand. 
If we simply take $M_m\sim\Mp$, which is the unique scale in this model,
the upper bound of the neutrino mass becomes $O(10^{-5}{\rm eV})$,
which is much smaller than the expected values $0.04 - 0.07$ eV
for the atmospheric neutrino anomaly.
Here we naively expect that in the effective term (\ref{M_N}), the factor
$\lambda^{l_i+l_j+2h_u}$ cannot be larger than 1, because terms
with negative total $U(1)_A$ charge $(l_i+l_j+2h_u<0)$ must be forbidden 
by the SUSY zero mechanism.
If we adopt the anomalous $U(1)_A$
gauge symmetry as the flavor symmetry that induces quark and lepton
masses, we have to explain why the mass
 of right-handed neutrinos 
is much smaller than that expected from the anomalous $U(1)_A$
charges, or to find a way to avoid the `SUSY zero' mechanism.
One might think that 
introducing a singlet whose VEV gives the mass of the right-handed 
neutrino can allow us to avoid this problem. Unfortunately, 
this solution does not 
work well if the $F$-flatness condition determines the VEV.
This is because, as we discuss in the next section,
the VEV of the singlet $S$ is generally determined by 
the anomalous $U(1)_A$ charge $s$ as
$\VEV{S}=\lambda^{-s}$, which does not improve the situation.
Of course, we could assume the right-handed neutrino scale
determined by some other conditions, for example, $D$-flatness
conditions, SUSY breaking terms, or  some dynamical mechanism.
In this paper, however, we examine more
appealing solutions to this problem. 
One of them is very simple. Note that even if we shift the anomalous
$U(1)_A$ charges $(q_i,u_i,d_i,l_i,e_i,n_i,h_u,h_d)$ to 
$(q_i+n,u_i+n,d_i+n,l_i+n,e_i+n,n_i+n,h_u-2n,h_d-2n)$,
the Dirac mass matrices of quarks and leptons remain unchanged.
On the other hand the right-handed neutrino masses become smaller 
by a factor of $\lambda^{2n}$ for positive $n$. Then the neutrino masses 
can be enhanced by a factor of $\lambda^{-2n}$. Note that even if the 
total charge $l_i+l_j+2h_u$ is negative, the term in Eq.~(\ref{M_N}) is 
allowed. This implies that 
the `SUSY zero' mechanism does not work in the effective interaction.
In the effective theory, 
which is obtained
by integrating heavy fields with positive anomalous $U(1)_A$ charges,
terms with negative total charge can be induced.  
It is easily shown
that the induced terms with negative total charge do not contribute to
the $F$-flatness conditions
if the heavy fields have vanishing VEVs. This observation is important
for the DT splitting models discussed in this paper, because
the SUSY zero mechanism plays an essential role to determine VEVs.
Note that integrating
heavy fields with masses of the Planck scale does not induce such terms,
because the total $U(1)_A$ charge of the mass term is zero.
This solution inevitably
leads to the negative charge of the Higgs field, which is required
also by the DT splitting mechanism proposed in this paper. 
For the other
solution, which can give a smaller mass to right-handed neutrino than 
that expected
from the anomalous $U(1)_A$ charge, 
it is essential that the right-handed neutrinos
have the charges of a gauge interaction.
We will return to this point in the next
section.

\section{Relation between VEVs and anomalous $U(1)_A$ charges 
and neutrino masses}
In this section, we discuss how VEVs are determined by the anomalous 
$U(1)_A$
quantum numbers.

First, the VEV of a gauge invariant operator with positive
anomalous $U(1)_A$ charge must vanish.
Otherwise, the SUSY zero mechanism does not work, since such a 
VEV can
compensate for the negative $U(1)_A$ charge of the term.
At this stage, such an undesired vacuum is not forbidden. However,
we show below that such a vacuum requires a VEV larger than the Planck
scale. If the cutoff is rigid and a VEV larger than the cutoff is not
allowed for some reason, then the condition for the SUSY zero 
mechanism is automatically satisfied.

Next we show that 
the VEV of a gauge invariant operator $O$ is generally 
determined by its $U(1)_A$ charge $o$ as $\VEV{O}=\lambda^{-o}$
if the $F$-flatness condition determines the VEV.
For simplicity, we examine this relation using singlet fields $Z_i$
with anomalous $U(1)_A$ charge $z_i$.
The general superpotential is written
\begin{eqnarray}
W&=&\sum_i \lambda^{z_i}Z_i+\sum_{i,j}\lambda^{z_i+z_j}Z_iZ_j+\cdots \\
&=&\sum_i \tilde Z_i+\sum_{i,j}\tilde Z_i\tilde Z_j+\cdots,
\end{eqnarray}
where $\tilde Z_i\equiv \lambda^{z_i}Z_i$.
The equations for the $F$-flatness of the $Z_i$ fields require
\begin{equation}
\lambda^{z_i}\left(1+\sum_j\tilde Z_j+\cdots\right)=0,
\end{equation}
which generally leads to solutions $\tilde Z_j\sim O(1)$. 
Note that at least one field of a term in the superpotential
must have positive or zero anomalous $U(1)_A$ charge. Otherwise
we cannot write down the term satisfying $U(1)_A$ gauge invariance.
As noted above, maintaining the SUSY zero feature requires 
that the VEV of a 
gauge invariant operator with positive anomalous $U(1)_A$ charge vanishes.
Therefore, with this requirement, 
usually it is sufficient to examine the $F$-flatness
of the gauge invariant operator with positive or zero anomalous $U(1)_A$ 
charges.
\footnote{
Note that the $F$-flatness condition of gauge variant fields with 
negative
charge can be important to determine the VEVs. This is because gauge 
variant fields with positive charge may have non-vanishing VEV.
}
(Therefore the $F$-flatness condition for $\Phi$ is automatically 
satisfied, because $\Phi$ has negative charge $-1$.)

Let us return to the problem involving the mass of the neutrino, which
is discussed in the previous section. From the above argument, it is 
shown that introducing 
a singlet field $S$ with non-zero VEV and the interaction 
$\lambda^{s+2n^c}SN_R^cN_R^c$
cannot improve the situation because the VEV of the
singlet is written $\VEV{S}=\lambda^{-s}$ if $F$-flatness 
conditions determine the VEV. 
Of course it is obvious that this problem can be avoided 
if the VEV is determined dynamically or by some other conditions, 
for example, $D$-flatness conditions
\cite{Ramond}.
However, we now propose another simple way to avoid this problem.

Let us introduce an additional gauge freedom\footnote{
A global symmetry can play the same role if the Nambu-Goldstone fields 
are phenomenologically allowed.
}
that transforms 
the right-handed neutrino fields non-trivially, for example, an 
additional $U(1)_X$ gauge symmetry or a gauge group larger than the 
standard
gauge group, like $SO(10)$. If the gauge variant field couples to the 
mass term of the right-handed neutrino and the VEV of the field 
breaks the additional gauge symmetry, then the coefficient can be 
changed. 
For example,  if we introduce 
additional singlets under the standard gauge group $\Theta (1,-6)$ and 
$\bar \Theta (-1,0)$, as well as 
the right-handed neutrinos $N_R^c(1,1)$, under the gauge group 
$U(1)_X\times U(1)_A$, then the VEV of the gauge invariant operator
$\VEV{\bar \Theta\Theta}$ is determined by the anomalous $U(1)_A$ 
charge $-6$ and is
given as $\VEV{\bar \Theta\Theta}=\lambda^6$. The mass term of the 
right-handed
neutrino is obtained from the term
$ \lambda^2(N_R^c(1,1)\bar \Theta (-1,0))^2$ with the VEV
$\VEV{\Theta (1,6)}=\VEV{\bar \Theta (-1,0)}\sim \lambda^3$, which is
required by the $D$-flatness condition of $U(1)_X$.
This mass term is of order $\lambda^8$, which is much smaller than
the naively expected value $\lambda^2$. 
The fact that the additional gauge freedom is required to obtain the 
correct
size of the mass of the right-handed neutrino implies that the 
GUT, if it exists, must have a rank greater than 4. The $SO(10)$ 
gauge group
is one of the most promising possibility, because it also unifies one 
generation of 
quarks and leptons, including the right-handed neutrino, in a single 
multiplet (spinor)
$\Psi$.
Actually if we adopt the $SO(10)$ gauge group, which is broken by the
VEV of the spinor $\VEV{C}=\VEV{\bar C}\sim \lambda^{-(c+\bar c)/2}$, 
the mass term of the right-handed neutrino is given from the term
$\lambda^{2(\psi+\bar c)}(\Psi\bar C)^2$. The mass 
$\lambda^{2\psi+\bar c-c}$ can be smaller than the naively expected 
value 
$\lambda^{2\psi}$. The model proposed by Bando and Kugo
\cite{Bando} has
such a structure in $E_6$ unification.

Such a solution, employing a larger unification group, is 
attractive, but GUT generally suffers from the DT splitting
problem.  
In the next section we show that the DT splitting is naturally
realized in $SO(10)$ unification using the anomalous $U(1)_A$ 
gauge symmetry.

\section{Doublet-triplet splitting with anomalous $U(1)_A$ gauge 
symmetry}
In the previous section, we emphasized that introducing the 
$SO(10)$ grand unified group or a larger group can naturally explain 
the mass scale of the right-handed neutrino. However, if we proceed 
to the unified theory with a simple group, we have to solve the 
doublet-triplet splitting problem. In this section, we propose 
an $SO(10)$ unified model that naturally realizes the 
doublet-triplet splitting. 

The content of the Higgs sector with $SO(10)\times U(1)_A$ gauge 
symmetry 
is given in Table I, where the symbols $\pm$ denote a parity quantum 
numbers.

\vspace{0.5cm}

\begin{center}
Table I. The lowercase letters represent the anomalous $U(1)_A$ charges.
\begin{eqnarray}
{\bf 45}&:& A(a=-2,-), \quad A^\prime (a^\prime=6,-)  \nonumber \\
{\bf 16}&:& C(c=-1,+), \quad C^\prime (c^\prime=8,-) \nonumber \\
{\bf \overline{16}}&:& \bar C(\bar c=-5,+), \quad 
                  \bar C^\prime (\bar c^\prime=4,-) \nonumber \\
{\bf 10}&:& H(h=-2,+), \quad H^\prime (h^\prime=4,-) \nonumber \\
{\bf 1}&:&  Z(z=-3,-), \quad \bar Z(\bar z=-3,-), \quad S(s=5,+) \nonumber
\end{eqnarray}
\end{center}
Here we have listed typical values of the anomalous
$U(1)_A$ charges. 
Among these fields, $A$, $C$, $\bar C$, $Z$ and $\bar Z$ are expected to
obtain non-vanishing
VEVs
around the GUT scale. Here, for simplicity we assume that the fields with 
positive $U(1)_A$ charges have vanishing VEVs, although we can give a more
rigorous argument for this.

Since the fields with non-vanishing VEVs have negative charges, only the 
$F$-flatness conditions of fields with positive charge must be counted
for determination of their VEVs. (Generally $c$ or $\bar c$ can be 
positive, 
although we are now considering $c=-1$ and $\bar c=-5$,
because it is sufficient for maintaining SUSY zero mechanism that 
the charge $c+\bar c$ 
of the gauge invariant operator $\bar CC$ become non-positive. 
The following argument does not change significantly if $c$ or $\bar c$
is positive.) 
Moreover, we have only to take account of
the terms in the superpotential which contain only one field with 
positive 
charge. This is because the terms with more positive charge fields 
do not 
contribute to the $F$-flatness conditions, since the positive fields 
are
assumed to have zero VEV.
Therefore, in general, the superpotential required by determination 
of the
VEVs can be written as
\begin{equation}
W=W_{H^\prime}+ W_{A^\prime} + W_S + W_{C^\prime}+W_{\bar C^\prime}.
\end{equation}
Here $W_X$ denotes the terms linear in the $X$ field, which has 
positive anomalous $U(1)_A$ charge. Note, however, that terms 
including two
fields with positive charge like 
$\lambda^{2h^\prime}H^\prime H^\prime$
give contributions to the mass terms but not to the VEVs.
$W_{A^\prime}$ can realize 
the DW form for the VEV
of $A$, $\VEV{A}=i\tau_2\times {\rm diag}(v,v,v,0,0)$, which is
proportional to the generator $B-L$.
Such a VEV of $A$ gives a super heavy mass to the color
triplets in $H$ and $H^\prime$ through the $W_{H^\prime}=HAH^\prime$ term, 
while keeping the weak doublets massless. 
This implies that the $F$-flatness condition of $H^\prime$ causes a 
vanishing VEV of the colored Higgs in $H$, but not the VEV of the 
doublet Higgs in $H$. 
The mass term 
$\lambda^{2h^\prime}(H^\prime)^2$ 
imparts a mass
$\sim \lambda^{2h^\prime}$ on the extra doublet in $H^\prime$. Therefore
it is realized that only one pair of doublet Higgs in $H$ becomes 
massless.

We now discuss the determination of the VEVs.
For determination of the VEVs, it is sufficient to take account of the
superpotential terms, which include only fields with non-zero VEVs,
except one field with vanishing VEV.
%$A^\prime$. 
If $-3a\leq a^\prime < -5a$,
the superpotential $W_{A^\prime}$ is in general
written as
\begin{equation}
W_{A^\prime}=\lambda^{a^\prime+a}\alpha A^\prime A+\lambda^{a^\prime+3a}(
\beta(A^\prime A)_{\bf 1}(A^2)_{\bf 1}
+\gamma(A^\prime A)_{\bf 54}(A^2)_{\bf 54}),
\end{equation}
where the suffixes {\bf 1} and {\bf 54} indicate the representation 
of the composite
operators under the $SO(10)$ gauge symmetry, and $\alpha$, $\beta$ and 
$\gamma$ are parameters of order 1. Here we assume 
$a+a^\prime+c+\bar c<0$
to forbid the term $\bar C A^\prime A C$, which destabilizes the 
DW form of the VEV $\VEV{A}$. 
If we take 
$\VEV{A}=i\tau_2\times {\rm diag}(x_1,x_2,x_3,x_4,x_5)$, the $F$-flatness
of the $A^\prime$ field requires
$x_i(\alpha\lambda^{-2a}+2(\beta-\gamma)(\sum_j x_j^2)+\gamma x_i^2)=0$, 
which gives only two solutions $x_i^2=0$, 
$\frac{\alpha}{(2N-1)\gamma-2N\beta}\lambda^{-2a}$. 
Here $N=1-5$ is the number of $x_i \neq 0$ solutions.
The DW form is obtained when $N=3$.
Note that the higher terms $A^\prime A^{2L+1}$ $(L>1)$ are forbidden by
the SUSY zero mechanism. If they are allowed, 
the number of possible VEVs other than the DW form
becomes larger, and thus it becomes less natural to obtain the DW form. 
This is a
critical point of this mechanism, and the anomalous $U(1)_A$ gauge 
symmetry
plays an essential role to forbid the undesired terms.
It is also interesting that the scale of the VEV is automatically
determined by the anomalous $U(1)_A$ charge of $A$, as noted in the 
previous section.

Next we discuss the $F$-flatness condition of $S$, which determines
the scale of the VEV $\VEV{\bar C C}$. 
$W_S$, which is linear in the $S$ field, is given by
\begin{equation}
W_S=\lambda^{s+c+\bar c}S\left((\bar CC)+\lambda^{-(c+\bar c)}
+\sum_k\lambda^{-(c+\bar c)+2ka}A^{2k}\right)
\end{equation}
if 
$s\geq -(c+\bar c)$.
Then the $F$-flatness condition of $S$ implies $\VEV{\bar CC}\sim 
\lambda^{-(c+\bar c)}$, and the $D$-flatness condition requires 
$|\VEV{C}|=|\VEV{\bar C}|\sim \lambda^{-(c+\bar c)/2}$.
The scale of the VEV is determined only by the charges of 
$C$ and $\bar C$ again.
If we take $c+\bar c=-6$, then we obtain the VEVs of the fields 
$\bar C$ and $\bar C$
as 
$\lambda^3$, which differ from the expected values $\lambda^{-c}$ 
and
$\lambda^{-\bar c}$ if $c\neq \bar c$.
Note that a composite operator with positive anomalous $U(1)_A$ charge
larger than $-(c+\bar c)-1$ may play the same role as the singlet $S$ if 
such a composite operator exists. (In the above example, there is no 
such composite operator.)

Finally, we discuss the $F$-flatness of $C^\prime$ and 
$\bar C^\prime$,
which realizes the alignment of the VEVs $\VEV{C}$ and $\VEV{\bar C}$
and imparts masses on the PNG fields. 
This simple mechanism
was proposed by Barr and Raby
\cite{BarrRaby}.
We can easily assign anomalous $U(1)_A$ charges which allow the 
following superpotential:
\begin{eqnarray}
W_{C^\prime}&=&
       \bar C(\lambda^{\bar c^\prime +c+a}A
       +\lambda^{\bar c^\prime +c+\bar z}\bar Z)C^\prime, \\
W_{\bar C^\prime}&=&
       \bar C^\prime(\lambda^{\bar c^\prime +c+a} A
       +\lambda^{\bar c^\prime +c+z}Z)C.
\end{eqnarray}
The $F$-flatness conditions $F_{C^\prime}=F_{\bar C^\prime}=0$ give
$(\lambda^{a-z} A+Z)C=\bar C(\lambda^{a-\bar z} A+\bar Z)=0$. 
Recall that the VEV of $A$ is 
proportional to the $B-L$ generator $Q_{B-L}$ as 
$\VEV{A}=\frac{3}{2}vQ_{B-L}$.
Also $C$, ${\bf 16}$, is decomposed into 
$({\bf 3},{\bf 2},{\bf 1})_{1/3}$, 
$({\bf \bar 3},{\bf 1},{\bf 2})_{-1/3}$, 
$({\bf 1},{\bf 2},{\bf 1})_{-1}$ and $({\bf 1},{\bf 1},{\bf 2})_{1}$ 
under
$SU(3)_C\times SU(2)_L\times SU(2)_R\times U(1)_{B-L}$.
Since $\VEV{\bar CC}\neq 0$, 
not all components in the spinor $C$ vanish. 
Then $Z$ is fixed to be $Z\sim -\frac{3}{2}\lambda v Q_{B-L}^0$, 
where $Q_{B-L}^0$ is
the $B-L$ charge of the component field in $C$, which has non-vanishing VEV. 
It is interesting
that no other component fields can have non-vanishing VEVs 
because of the $F$-flatness
conditions. If  the $({\bf 1},{\bf 1},{\bf 2})_{\bf 1}$ field obtains
a non-zero VEV (therefore, $\VEV{Z}\sim -\frac{3}{2}\lambda v$), then the 
gauge group 
$SU(3)_C\times SU(2)_L\times SU(2)_R\times U(1)_{B-L}$ is broken to the 
standard gauge group. Once the direction of the VEV $\VEV{C}$ is 
determined, the VEV $\VEV{\bar C}$ must have the same direction 
because of the $D$-flatness
condition. Therefore, $\VEV{\bar Z}\sim -\frac{3}{2}\lambda v$.
Thus, all VEVs have now been fixed.

Next we examine the mass spectrum. Since for the mass terms, we must
take account of not only the above terms but also
the terms that contain two fields with vanishing
VEVs. 

Considering the additional mass term
$\lambda^{2h^\prime} H^\prime H^\prime$, we write the mass matrix of 
the Higgs fields $H$ and $H^\prime$, which are decomposed from {\bf 5} 
and 
${\bf \bar 5}$ of $SU(5)$, as
\begin{equation}
({\bf 5}_H, {\bf 5}_{H^\prime})
\left(\begin{array}{cc} 0 & \lambda^{h+h^\prime +a}\VEV{A} \\
                     \lambda^{h+h^\prime +a}\VEV{A} & \lambda^{2h^\prime}
      \end{array}\right)
\left(\begin{array}{c} {\bf \bar 5}_H \\ {\bf \bar 5}_{H^\prime}
\end{array}\right).
\end{equation}
The colored Higgs obtain their masses of order 
$\lambda^{h+h^\prime+a}\VEV{A}\sim \lambda^{h+h^\prime}$.
Since in general $\lambda^{h+h^\prime}>\lambda^{2h^\prime}$,
the proton decay is naturally suppressed. The effective colored
Higgs mass is estimated as
$(\lambda^{h+h^\prime})^2/\lambda^{2h^\prime}=\lambda^{2h}$, 
which is larger than the Planck scale, because
$h<0$.
One pair of the doublet Higgs is massless,
while another pair of doublet Higgs acquires a mass of order 
$\lambda^{2h^\prime}$, which is $\sim\lambda^8\sim 5\times 10^{12}$ GeV
in the typical $U(1)_A$ assignment in Table I. 
The DW mechanism works well, although we have to examine the effect of the 
rather
light additional Higgs.

Next we examine the mass matrices for the representations 
$I=Q,U^c$ and $E^c$,
which are contained in the {\bf 10} of $SU(5)$.
Like the superpotential previously discussed, the additional terms
$\lambda^{2a^\prime}A^\prime A^\prime$, 
$\lambda^{c^\prime+\bar c^\prime}\bar C^\prime C^\prime$,
$\lambda^{c^\prime+a^\prime+\bar c} \bar CA^\prime C^\prime$ and
$\lambda^{\bar c^\prime+a^\prime+c} \bar C^\prime A^\prime C$
must be taken into account.
The mass matrices are written as $4\times 4$ matrices,
\begin{equation}
\left(\bar I_A, \bar I_{A^\prime},\bar I_{\bar C},\bar I_{\bar C^\prime}
\right)\left(
\begin{array}{cccc}
0& \lambda^{a^\prime+a} \alpha_I & 0  & 
\frac{\lambda^{\bar c+c^\prime+a}}{\sqrt{2}}\VEV{\bar C} \\
\lambda^{a+a^\prime} \alpha_I & \lambda^{2a^\prime} & 0 & 
\frac{\lambda^{\bar c+c^\prime+a^\prime}}{\sqrt{2}}\VEV{\bar C} \\
0 & 0 & 0 & \lambda^{\bar c+c^\prime+a}\beta_Iv \\
\frac{\lambda^{c+\bar c^\prime+a}}{\sqrt{2}}\VEV{C} &
\frac{\lambda^{c+\bar c^\prime+a^\prime}}{\sqrt{2}}\VEV{C} &
\lambda^{c+\bar c^\prime+a}\beta_Iv & \lambda^{c^\prime+\bar c^\prime}
\end{array}
\right)\left(
\begin{array}{c}
I_A \\ I_{A^\prime} \\ I_C \\ I_{C^\prime}
\end{array}
\right),
\label{mass10}
\end{equation}
where $\alpha_I$ vanishes for $I=Q$ and $U^c$ because
these are Nambu-Goldstone modes, but $\alpha_{E^c}\neq 0$.
On the other hand, $\beta_I=\frac{3}{2}((B-L)_I-1)$; that is, $\beta_Q=-1$, 
$\beta_{U^c}=-2$ and $\beta_{E^c}=0$. 
Thus for each $I$, the $4\times 4$ matrix has one 
vanishing eigenvalue, which corresponds to the Nambu-Goldstone mode eaten
by the Higgs mechanism. The mass spectrum of the remaining three
modes is ($\lambda^{c+\bar c^\prime+a}v$, $\lambda^{c^\prime+\bar c+a}v$,
$\lambda^{2a^\prime}$) for the color-triplet modes $Q$ and $U^c$, and 
($\lambda^{a+a^\prime}$, 
$\lambda^{a+a^\prime}$,
$\lambda^{c^\prime+\bar c^\prime}$) or 
($\lambda^{c+\bar c^\prime+a}\VEV{C}$, 
$\lambda^{c^\prime+\bar c+a}\VEV{\bar C}$,
$\lambda^{2a^\prime}$) for the color-singlet modes $E^c$. 
(These are dependent on the anomalous $U(1)_A$ charges.)
If we use typical anomalous $U(1)_A$ charges, as listed in the previous 
table, then the light modes
are $Q$, $U^c$, $E^c$ and their hermitian conjugate fields, which are
contained in a pair of ${\bf 10}$ and ${\bf \overline{10}}$ of $SU(5)$,
with a mass of order $\lambda^{12}\sim 10^{10}$ GeV.
Though in principle the mass of the color-triplet fields and
that of the color-singlet field are determined independently, it is
interesting that all the fields in a single multiplet {\bf 10} of $SU(5)$
become light together. 
This fact makes us expect that the
success of the gauge coupling unification may not be drastically changed
with these light modes.

If we simply omit the rows and columns of $A$ and $A^\prime$ in 
Eq.~(\ref{mass10}), 
then we obtain $2\times 2$
mass matrices, which are for the representations $D^c$ and $L$ and their
conjugates. Since $\beta_{D^c}=-2$ and $\beta_L=-3$, the color 
triplets acquire masses $2\lambda^{\bar c+c^\prime}$ and 
$2\lambda^{c+\bar c^\prime}$, while the weak doublets acquire masses  
$3\lambda^{\bar c+c^\prime}$ and 
$3\lambda^{c+\bar c^\prime}$.

The adjoint fields $A$ and $A^\prime$ contain
two $({\bf 8},{\bf 1})_{0}$ and two $({\bf 1},{\bf 3})_{0}$
of the standard gauge group,
which acquire mass $\lambda^{a^\prime+a}$. Moreover, they contain
two pairs of $({\bf 3},{\bf 2})_{-5/6}$+h.c. One of these is
eaten by Higgs mechanism, but another pair has a rather light mass of
$\lambda^{2a^\prime}$, which may destroy the coincidence
of the running gauge couplings.

Once we determine the anomalous $U(1)_A$ charges, the mass spectrum
of all fields is determined, and hence we can calculate the Weinberg angle.
However, since the estimation is strongly dependent on the assignment
of the anomalous $U(1)_A$ charges, as shown in the above argument, and
on the details of 
the DT splitting sector and the
matter sector, we do not discuss it further here.

There are several terms which must be forbidden for the stability of the 
DW mechanism. For example, $H^2$, $HZH^\prime$ and $H\bar Z H^\prime$
induce a large mass of the doublet Higgs, 
and the term $\bar CA^\prime A C$ would destabilize the DW form of 
$\VEV{A}$.
We can easily forbid these terms using the SUSY zero mechanism.
For example, if we choose
$h<0$, then $H^2$ is forbidden, and if we choose $\bar c+c+a+a^\prime<0$, 
then
$\bar CA^\prime A C$ is forbidden. (It is interesting that the negative 
$U(1)_A$ charge $h$, which is required for the DT splitting, enhances
the left-handed neutrino masses, as discussed in section 2.) 
Once these dangerous terms are forbidden
by the SUSY zero mechanism, higher-dimensional terms which also become
dangerous;
for example, 
$\bar CA^\prime A^3 C$ and $\bar CA^\prime C\bar CA C$ are automatically
forbidden, since only gauge invariant operators with negative charge
can have non-vanishing VEVs. This is also an attractive point of our 
scenario. 
Actually, 
the symmetry discussed in Ref.\cite{Barr} does not forbid the dangerous
term $(\bar CA C)^2$, which destabilizes the DW form of $\VEV{A}$.

In this section, we have proposed an natural DT splitting mechanism 
in which the anomalous $U(1)_A$ gauge symmetry play a critical role, 
and the VEVs and mass spectrum are automatically determined by the 
anomalous $U(1)_A$ charges. 
In the next section, we examine the simplest model with this DT 
splitting 
mechanism, which gives realistic mass matrices of quarks and leptons.

\section{The simplest model}
In this section, we examine the simplest model to demonstrate 
how to determine
everything from the anomalous $U(1)_A$ charges.

In addition to the Higgs sector in Table.I, we introduce only three 
${\bf 16}$ representations $\Psi_i$
with anomalous $U(1)_A$ charges $(\psi_1=n+3,\psi_2=n+2,\psi_3=n)$ 
and one ${\bf 10}$ field
$T$ with
charge $t$ as the matter content. These matter fields are assigned
odd R-parity, while those of the Higgs sector are assigned even 
R-parity.
Such an assignment of R-parity guarantees that the argument regarding
VEVs 
in the previous section does not change if these matter fields have 
vanishing VEVs.
We can give an argument to determine the allowed region of the 
anomalous $U(1)_A$ charges 
to obtain
desired terms while forbidding dangerous terms. 
Though this is a straightforward
argument, we do not give it here. Instead, 
we give a set of anomalous $U(1)_A$
charges with which all conditions are satisfied and a novel neutrino
mass matrix is obtained:
$n=3, t=4, h=-6, h^\prime=8, c=-4, \bar c=-1, c^\prime=4, 
\bar c^\prime=7, s=5$.
Then the mass term of ${\bf 5}$ and ${\bf \bar 5}$ of $SU(5)$
is written as
\begin{equation}
{\bf 5}_T ( \lambda^6 \VEV{C}, \lambda^5 \VEV{C}, \lambda^3\VEV{C}, 
\lambda^8)
\left(
\begin{array}{c}  {\bf \bar 5}_{\Psi1} \\ {\bf \bar 5}_{\Psi2} \\ 
{\bf \bar 5}_{\Psi3}
 \\ {\bf \bar 5_T}
\end{array}
\right).
\end{equation}
Since $\VEV{\bar C}=\VEV{C}\sim \lambda^{5/2}$, because $c+\bar c=-5$, 
the massive mode ${\bf \bar 5}_M$, the partner of $5_T$, is given by
\begin{equation}
{\bf \bar 5}_M \sim {\bf \bar 5}_{\Psi3}+\lambda^{5/2}{\bf \bar 5}_T.
\end{equation} 
Therefore the three massless modes 
$({\bf \bar 5}_1, {\bf \bar 5}_2, {\bf \bar 5}_3) $ are
written $({\bf \bar 5}_{\Psi1}, 
{\bf \bar 5}_T+ \lambda^{\frac{5}{2}}{\bf \bar 5}_{\Psi3},
{\bf \bar 5}_{\Psi2})$. The Dirac mass matrices for quarks and leptons 
can be 
obtained from the interaction
\begin{equation}
\lambda^{\psi_i+\psi_j+h}\Psi_i\Psi_jH.
\label{Yukawa}
\end{equation}
The mass matrices for the up quark sector and the down quark sector are
\begin{equation}
M_u=\left(
\begin{array}{ccc}
\lambda^6 & \lambda^5 & \lambda^3 \\
\lambda^5 & \lambda^4 & \lambda^2 \\
\lambda^3 & \lambda^2 & 1    
\end{array}
\right)\VEV{H_u},\quad
M_d=\lambda^2\left(
\begin{array}{ccc}
\lambda^4 & \lambda^{7/2} & \lambda^3 \\
\lambda^3 & \lambda^{5/2} & \lambda^2 \\
\lambda^1 & \lambda^{1/2} & 1    
\end{array}
\right)\VEV{H_d}.
\end{equation}
Note that the Yukawa couplings for 
${\bf \bar 5}_2\sim {\bf \bar 5}_T+\lambda^{5/2}{\bf \bar 5}_{\Psi 3}$
are obtained only through the Yukawa couplings for the component 
${\bf \bar 5}_{\Psi 3}$,
because we have no Yukawa couplings for $T$.
We can estimate the CKM matrix from these quark matrices as
\begin{equation}
U_{CKM}=
\left(
\begin{array}{ccc}
1 & \lambda &  \lambda^3 \\
\lambda & 1 & \lambda^2 \\
\lambda^3 & \lambda^2 & 1
\end{array}
\right),
\end{equation}
which is consistent with the experimental value if we choose 
$\lambda\sim 0.2$.
Since the ratio of the Yukawa couplings of top and bottom quarks is 
$\lambda^2$,
a small value of $\tan \beta\equiv \VEV{H_u}/\VEV{H_d} \sim O(1)$ is 
predicted by these mass matrices.
The Yukawa matrix for the charged lepton sector is the same as the transpose 
of $M_d$ at this stage, except for an overall factor $\eta$ induced by the 
renormalization group effect. 
The mass matrix for the Dirac mass of neutrinos is given by
\begin{equation}
M_D=\lambda^2\left(
\begin{array}{ccc}
\lambda^4 & \lambda^3 & \lambda \\
\lambda^{7/2} & \lambda^{5/2} & \lambda^{1/2} \\
\lambda^3   & \lambda^2         & 1 
\end{array}
\right)\VEV{H_u}\eta.
\end{equation}

The right-handed neutrino masses come from the interaction
\begin{equation}
\lambda^{\psi_i+\psi_j+2\bar c}\Psi_i\Psi_j\bar C\bar C
\end{equation}
as
\begin{equation}
M_R=\lambda^{\psi_i+\psi_j+2\bar c}\VEV{\bar C}^2=\lambda^9\left(
\begin{array}{ccc}
\lambda^6 & \lambda^5 & \lambda^3 \\
\lambda^5 & \lambda^4 & \lambda^2 \\
\lambda^3   & \lambda^2         & 1
\end{array}
\right).
\end{equation}
Therefore we can estimate the neutrino mass matrix:
\begin{equation}
M_\nu=M_DM_R^{-1}M_D^T=\lambda^{-5}\left(
\begin{array}{ccc}
\lambda^2 & \lambda^{3/2} & \lambda \\
\lambda^{3/2} & \lambda & \lambda^{1/2} \\
\lambda   & \lambda^{1/2}         & 1 
\end{array}
\right)\VEV{H_u}^2\eta^2.
\end{equation}
Note that the overall factor $\lambda^{-5}$ has
negative power, which
can be induced by the effects discussed in sections 2 and 3.
From these mass matrices in the lepton sector the MNS
matrix is obtained as
\begin{equation}
U_{MNS}=
\left(
\begin{array}{ccc}
1 & \lambda^{1/2} &  \lambda \\
\lambda^{1/2} & 1 & \lambda^{1/2} \\
\lambda & \lambda^{1/2} & 1
\end{array}
\right).
\end{equation}
This gives bi-maximal mixing angles for the neutrino sector,
because $\lambda^{1/2}\sim 0.5$.\footnote{
After submitting this paper, we noticed the reference 
\cite{Matsuoka} in which this neutrino mass structure has been 
discussed with the semi-simple unified group $SU(6)\times SU(2)_R$. 
}
We then obtain the prediction 
$m_{\nu_\mu}/m_{\nu_\tau}\sim \lambda$, which is consistent with
the experimental data: 
$1.6\times 10^{-3} ({\rm eV})^2\leq \Delta m_{\rm atm}^2\leq 4
\times 10^{-3}
({\rm eV})^2$
and $2\times 10^{-5} ({\rm eV})^2\leq \Delta m_{\rm solar}^2\leq 1
\times 10^{-4}
({\rm eV})^2$.
The relation $V_{e3}\sim \lambda$ is also an interesting 
prediction from this 
matrix, though CHOOZ gives a restrictive upper limit $V_{e3}\leq 0.15$
\cite{CHOOZ}.
The neutrino mass is given by
$m_{\nu_\tau}\sim \lambda^{-5}\VEV{H_u}^2\eta^2/\Mp\sim 
m_{\nu_\mu}/\lambda
\sim m_{\nu_e}/\lambda^2$. If we take $\VEV{H_u}\eta=100$ GeV, 
$\Mp\sim 10^{18}$ GeV and $\lambda=0.2$, then we get 
$m_{\nu_\tau}\sim 3\times 10^{-2}$ eV, $m_{\nu_\mu}\sim 6\times 
10^{-3}$ eV
and $m_{\nu_e}\sim 1\times 10^{-3}$ eV. 
It is surprising that such a rough approximation gives
values in good agreement with
the experimental values from the atmospheric neutrino and large 
mixing angle (LMA) MSW solution of the solar neutrino problem
\cite{MSW}. This LMA solution for the solar neutrino problem
gives the best fit to the present experimental data
\cite{Valle}.

In addition to Eq.~(\ref{Yukawa}), the interactions
\begin{equation}
\lambda^{\psi_i+\psi_j+2a+h}\Psi_iA^2\Psi_jH
\end{equation}
also contribute to the Yukawa couplings. Here $A$ is squared because
it has odd parity.
Since $A$ is proportional to the generator of $B-L$,
the contribution to the lepton Yukawa coupling is nine times larger
than that to quark Yukawa coupling, which can change the unrealistic
prediction $m_\mu=m_s$ at the GUT scale. 
Since the prediction $m_s/m_b\sim \lambda^{5/2}$ at the GUT scale is 
consistent with experiment, 
the enhancement factor $2\sim 3$ of $m_\mu$ can improve the situation.
Note that the additional terms contribute mainly in the lepton
sector.
If we set $a=-2$,
the additional matrices are
\begin{eqnarray}
\frac{\Delta M_u}{\VEV{H_u}}&=&
\frac{v^2}{4}\left(
\begin{array}{ccc}
\lambda^2 & \lambda & 0 \\
\lambda & 1 & 0 \\
0 & 0 & 0
\end{array}
\right)
,\quad
\frac{\Delta M_d}{\VEV{H_d}}=
\frac{v^2}{4}\left(
\begin{array}{ccc}
\lambda^2 & 0 & \lambda \\
\lambda & 0 & 1 \\
0 & 0 & 0 
\end{array}
\right)
,
\\
\frac{\Delta M_e}{\VEV{H_d}}&=&
\frac{9v^2}{4}\left(
\begin{array}{ccc}
\lambda^2 & \lambda & 0 \\
0 & 0 & 0 \\
\lambda & 1 & 0 
\end{array}
\right).
\end{eqnarray}
It is interesting that this modification essentially changes the 
eigenvalues of
only the first and second generation. Therefore it is natural to expect 
that a realistic mass pattern can be obtained by this modification.
This is one of the largest motivations to choose $a=-2$.
Note that this charge assignment also determines the scale 
$\VEV{A}\sim \lambda^2$.
It is suggestive that the fact that the $SO(10)$ breaking scale is 
slightly
smaller than the Planck scale is correlated with the discrepancy between
the naive prediction of the ratio $m_\mu/m_s$ from the unification and 
the experimental value. 
It is also interesting that the SUSY zero mechanism plays an essential 
role again. 
When $z, \bar z \geq -4$, the terms 
$\lambda^{\psi_i+\psi_j+a+z+h}Z\Psi_iA\Psi_jH+
\lambda^{\psi_i+\psi_j+2z+h}Z^2\Psi_i\Psi_jH
$ also contribute to the fermion mass matrices, though only to the first
generation.

Proton decay mediated by the colored Higgs is strongly suppressed
in this model. As mentioned in the previous section, the effective 
mass of
the colored Higgs is of order $\lambda^{2h}\sim \lambda^{-12}$, which is 
much larger than the Planck scale. 
Proton decay is also induced by the non-renormalizable
term
\begin{equation} 
\lambda^{\psi_i+\psi_j+\psi_k+\psi_l}\Psi_i\Psi_j\Psi_k\Psi_l,
\end{equation}
which is also strongly suppressed.

Unfortunately, in this model we obtain an extra light doublet Higgs 
with mass of order
$\lambda^{2h^\prime}\sim \lambda^{16}$ and extra fields 
$({\bf 3},{\bf 2})_{-5/6}$ + h.c. with mass of
order $\lambda^{2a^\prime}\sim \lambda^{12}$, which may destroy 
 gauge coupling unification.
Actually, a rough approximation shows that the meeting point of the gauge 
couplings of $SU(3)_C$ and $SU(2)_L$ is too low to maintain the proton
stability for any $U(1)_A$ charge assignment. 
However, since
the mass spectrum is strongly dependent on 
the details of DT splitting models and the matter sector,
we expect that for a certain charge assignment of a certain DT splitting
model and the matter sector, 
the coupling unification is recovered. In other words, the requirement
of coupling unification represents a strong constraint on these models.

\section{Discussion}
In this paper, we have examined a DT splitting model and emphasized that 
the anomalous $U(1)_A$ gauge symmetry plays
an essential role to realize DT splitting by the DW mechanism. 
This statement is generally true.
Actually, we can 
make various types of DT splitting models in which the anomalous $U(1)_A$ 
gauge symmetry plays the role discussed in this paper. 
For example, if we exchange the parity between $C^\prime$
and $\bar C$, the term $CC^\prime H$ is allowed. After obtaining the 
VEV
of the $C$ field, the massless doublet Higgs becomes a linear combination of
${\bf \bar 5}_H$ and ${\bf \bar 5}_{C^\prime}$ of $SU(5)$.
This may give richer structure to 
the quark and lepton matrices, though a dangerous term 
$CA^\prime\bar C$ must be forbidden by the SUSY zero mechanism.
We can introduce an additional 
Higgs pair, $F:{\bf 16}$ and $\bar F:{\bf \overline{16}}$, to obtain a 
massless doublet Higgs
that is linear combination of ${\bf \bar 5}_H$ and ${\bf \bar 5}_{F}$. 
Yet another way to modify the DT splitting model is to introduce an additional 
adjoint field ${\bf 45}$ $A_+(a_+,+)$ with $a_+<0$.
Then the mass spectrum of light modes is quite
different from that of the model studied in this paper, because
of the term $A^\prime A_+ A$.
Moreover, we have assumed that the anomalous $U(1)_A$ charges take integer
values, 
but 
in principle, they can take rational values as in Ref.\cite{Shafi}.
We have not carefully examined 
all these
modified DT
splitting models.
We will examine various possibilities in the future.
The condition for gauge coupling unification must be a useful guide
to select these models.

In principle, we may adopt an anomaly-free $U(1)_A$ gauge symmetry and 
the 
F-I $D$-term instead of the anomalous $U(1)_A$ gauge symmetry, 
though it seems to
be difficult to find a consistent $U(1)_A$ charge assignment. 
Moreover, since we have no reason
to choose the scale of the F-I $D$-term to be less than the Planck scale,
we think it more
natural to adopt the anomalous $U(1)_A$ gauge symmetry.

Finally, we discuss SUSY breaking. Since the
anomalous $U(1)_A$ charges should depend on the flavor to produce
a hierarchy of Yukawa couplings, generally
non-degenerate scalar fermion masses are induced through the anomalous 
$U(1)_A$
$D$-term. The large SUSY breaking scale allows us avoid the flavor changing
neutral current (FCNC) problem \cite{DG,CKN}, but in the present scenario
it does not work, because the anomalous $U(1)_A$ charge of the Higgs 
$H$ must be
 negative to forbid the Higgs mass term at tree level.
Therefore the anomalous $U(1)_A$ $D$-term contribution, which is 
dependent on the flavor, must be dominated by other flavor-independent 
contribution to
the sfermion mass terms. In principle, it is possible, for example, that 
the $F$-term
of the dilation field dominates the dangerous $D$-term contribution. 
In fact, Arkani-Hamed, Dine and Martin \cite{ADM} pointed out that the 
$F$-term contribution of the dilaton field can be larger than
the anomalous $U(1)_A$ $D$-term contribution, depending on how the
dilaton is stabilized, even in the case that the anomalous $U(1)_A$ 
gauge symmetry triggers SUSY
breaking\cite{BD,DP}.
\footnote{
In our case, it is difficult for the anomalous $U(1)_A$ gauge symmetry 
to trigger
SUSY breaking, because we have many fields with negative charge in addition
to the field $\Phi$.}
It is interesting that the lepton flavor violation 
process can be seen in this scenario
\cite{kurosawa}. Since the FCNC process introduces severe constraints on
the ratio of the $D$-term contribution and the flavor independent 
contribution, it is valuable to examine the condition for which the 
constraints become weaker. If the first generation of 
${\bf \bar 5}$ of $SU(5)$ has the same charge as 
the second generation, the constraint becomes
much weaker
\cite{kurosawa}. The condition for the model discussed in this paper 
is $\psi_1=t$. 
If we assign the anomalous $U(1)_A$ charge as
$n=5, t=8, a=-2, a^\prime=6, h=-10, h^\prime=12, c=-2, \bar c=-3, 
c^\prime=6, \bar c^\prime=5, z=\bar z=-3, s=5$, the above situation is 
realized, although the mass of an additional pair of Higgs becomes 
$\lambda^{24}$.
Even if the above effect is negligible, the lepton flavor violation 
process may be seen through the renormalization effect of 
the left-handed slepton masses
\cite{LFV}.

In subsequent papers
\cite{BKM}, it is shown that the DT splitting mechanism 
can be non-trivially incorporated into $E_6$ unification. It is 
interesting
that the mass matrices with bi-maximal mixing discussed in 
this paper appear again in the $E_6$ unified model. Moreover, the above
condition $\psi_1=t$, which makes the constraints from the FCNC process 
weaker,
is automatically satisfied.

\section{Conclusion}
In this paper, we have pointed out that, in order to realize the correct 
size of
the neutrino mass for the atmospheric neutrino anomaly with the anomalous 
$U(1)_A$ gauge symmetry,
it is natural to introduce a Higgs field with negative $U(1)_A$ charge 
and
a gauge group under which the right-handed neutrino
transforms non-trivially, for example, $SO(10)$, $E_6$, or extra 
$U(1)$.
Next we proposed an $SO(10)$ unified model in which DT splitting is 
naturally realized by the DW mechanism. 
The anomalous $U(1)_A$ gauge symmetry plays an essential role in the DT 
splitting.
Using this mechanism, we examined the simplest model in which realistic
mass matrices of quarks and leptons,
including the neutrino, can be determined by the anomalous $U(1)_A$ charges.
This model predicts bi-maximal mixing angles in the neutrino sector, a small 
value of $\tan \beta$, and the relation $V_{e3}\sim \lambda$. 
Proton stability is naturally 
realized.
It is interesting that once we fix
the anomalous $U(1)_A$ charges for all fields, 
the order of each parameter and 
scale is determined, except that of the SUSY breaking.

It is very suggestive that the anomalous $U(1)_A$ gauge symmetry 
motivated
by superstring theory plays a critical role in solving the two 
biggest 
problems in GUT, the fermion mass hierarchy problem and the 
doublet-triplet
splitting problem. This may be the first evidence for the 
validity of string theory from the phenomenological point of view.

\section{Acknowledgements}
We would like to thank M. Bando and T. Kugo for interesting discussions, 
careful reading of this manuscript, and useful comments. 
We also appreciate valuable comments from N. Haba and thank S. Yamashita
for correcting typos in this paper.
We also thank to the organizers of the workshop on neutrino in Kyoto, where
we obtained an opportunity to reconsider the mass of neutrino with anomalous
$U(1)_A$ gauge symmetry.


\begin{thebibliography}{99}
\bibitem{georgi}  H. Georgi and S.L. Glashow,  Phys. Rev. Lett.
                  {\bf 32} (1974) 438.
\bibitem{SUSYGUT} E. Witten, Nucl. Phys. B{\bf 188} (1981) 513;\\
                  S. Dimopoulos, S. Raby and F. Wilczek, Phys. Rev. D
                  {\bf 24} (1981) 1681;\\
                  D. Dimopoulos and H. Georgi,  Nucl. Phys. B {\bf 193}
                  (1981) 150;\\
                  N. Sakai, Z. Phys. C{\bf 11} (1981) 153.
\bibitem{SK}        Fukuda et al.(The Super-Kamiokande Collaboration), 
                    Phys. Lett. B{\bf 436} (1998) 33;
                     Phys. Rev. Lett. {\bf 81} (1998) 1562.
\bibitem{Sato}      J. Sato and T. Yanagida,  Phys.Lett. B {\bf 430} 
                    (1998) 127.
\bibitem{Nomura}    Y. Nomura and T. Yanagida,  Phys.Rev. D {\bf 59} 
                    (1999) 017303;\\
                    K.-I. Izawa, Kiichi Kurosawa, Yasunori Nomura, T. Yanagida
                     Phys. Rev. D{\bf 60} (1999) 115016.
\bibitem{Bando}  M. Bando and T. Kugo, Prog. Theor. Phys. {\bf 101}
                 (1999) 1313; \\
                 M. Bando, T. Kugo and K. Yoshioka,  Prog. Theor. Phys.
                 {\bf 104} (2000) 211.
\bibitem{Barr}   C.H. Albright, K.S. Babu and S.M. Barr,   Phys.Rev.Lett.
                 {\bf 81} (1998) 1167;\\
                 C.H. Albright and S.M. Barr,  Phys.Rev.Lett. {\bf 85}
                  (2000) 244;
                 Phys.Rev. D{\bf 62} (2000) 093008;
                 Phys.Lett. B{\bf 461} (1999) 218;
                  Phys.Lett. B{\bf 452} (1999) 287;
                 Phys.Rev. D{\bf 58} (1998) 013002.
\bibitem{Shafi} Q. Shafi and Z. Tavartkiladze,  Phys. Lett. B{\bf 487}
                 (2000) 145.
\bibitem{DTsplitting} E. Witten, Phys. Lett. B{\bf 105} (1981) 267;\\
                      A. Masiero, D.V. Nanopoulos, K. Tamvakis and T. Yanagida, 
                      Phys. Lett. {\bf 115} (1982) 380;\\
                      K. Inoue, A. Kakuto and T. Takano,
                      Prog. Theor. Phys. {\bf 75} (1986) 664;\\
                      E. Witten,  Nucl. Phys. B{\bf 258} (1985) 75;\\
                      T. Yanagida, Phys. Lett. B{\bf 344}  (1995) 211;\\
                      Y. Kawamura, hep-ph/0012125.
\bibitem{DW}       S. Dimopoulos and F. Wilczek, NSF-ITP-82-07;\\
                   M. Srednicki, Nucl. Phys. B{\bf 202} (1982) 327.
\bibitem{BarrRaby} S.M. Barr and S. Raby, Phys. Rev. Lett.
                   {\bf 79} (1997) 4748.
\bibitem{Chako}    Z. Chacko and R.N. Mohapatra,
                   Phys.Rev. D{\bf 59} (1999) 011702;
                   Phys.Rev.Lett. {\bf 82} (1999) 2836.
\bibitem{complicate} K.S. Babu and S.M. Barr, Phys. Rev. D{\bf 48}
                    (1993) 5354; ibid {\bf 50} (1994) 3529.
\bibitem{U(1)}    E. Witten,
                  Phys. Lett. B{\bf 149} (1984) 351;\\
                  M. Dine, N. Seiberg and E. Witten,
                  Nucl. Phys. B{\bf 289} (1987) 589;\\
                  J.J. Atick, L.J. Dixon and A. Sen,
                  Nucl. Phys. B{\bf 292} (1987) 109;\\
                  M. Dine, I. Ichinose and N. Seiberg,
                  Nucl. Phys. B{\bf 293} (1987) 253.
\bibitem{GS}      M. Green and J. Schwarz,
                  Phys. Lett. B{\bf 149} (1984) 117. 
\bibitem{FN}      C.D. Froggatt and H.B. Nielsen,
                  Nucl. Phys. B{\bf 147} (1979) 277.
\bibitem{Ibanez}  L. Ib\'a\~nez and G.G. Ross,
                  Phys. Lett. B{\bf 332} (1994) 100;\\
                  P. Bin\'etruy and P. Ramond,
                  Phys. Lett. B{\bf 350} (1995) 49;\\
                  E. Dudas, S. Pokorski and C.A. Savoy,
                  Phys. Lett. B{\bf 356} (1995) 45;\\
                  P. Bin\'etruy, S. Lavignac and P. Ramond,
                  Nucl. Phys. B{\bf 477} (1996) 353.
\bibitem{Ramond}  P. Bin\'etruy, S. Lavignac, S. Petcov and P. Ramond,
                  Nucl. Phys. B{\bf 496} (1997) 3.
\bibitem{Dreiner} H. Dreiner, G.K. Leontaris, S. Lola, G.G. Ross and C. Scheich,
                  Nucl. Phys. B{\bf 436} (1995) 461.
\bibitem{CKM}     M. Kobayashi and T. Maskawa, Prog. Theor. Phys.
                  {\bf 49} (1973) 652.
\bibitem{seasaw}  T. Yanagida, in {\it Proceedings of the Workshop on the 
                  Unified Theory and Baryon Number in the Universe}, eds.
                  O. Sawada and A. Sugamoto(KEK report 79-18,1979);\\
                  M. Gell-Mann, P. Ramond and R. Slansky, in {\it Supergravity},
                  eds. P.van Nieuwenhuizen and D.Z. Freedman (North Holland,
                  Amsterdam, 1979).
\bibitem{MNS}     Z. Maki, M. Nakagawa and S. Sakata,
                  Prog. Theor. Phys. {\bf 28} (1962) 870.
\bibitem{anarchy} L. Hall, H. Murayama and N. Weiner, Phys. Rev. Lett.
                  {\bf 84} (2000) 2572;\\
                  N. Haba and H. Murayama, Phys. Rev. D{\bf 63}
                   (2001) 053010.
\bibitem{Matsuoka} M. Matsuda and T. Matsuoka, Phys.Lett. B{\bf 499} (2001) 287.
\bibitem{CHOOZ}   The CHOOZ Collaboration, M. Appollonio et al.,
                  Phys. Lett. B{\bf 420}(1998) 397.
\bibitem{MSW}     L. Wolfenstein, Phys. Rev. D{\bf 17} (1978) 2369;\\
                  S.P. Mikheev and A. Smirnov,  Yad. Fiz. {\bf 42}
                  (1985) 1441; Nuovo Cim. {\bf 9C} (1986) 17.
\bibitem{Valle}   J.W.F. Valle, astro-ph/0104085; \\
                  M.C. Gonzalez-Garcia, M. Maltoni, C. Pena-Garay and 
                  J.W.F. Valle, Phys.Rev. D{\bf 63} (2001) 033005.
\bibitem{DG}      S. Dimopoulos and G.F. Giudice, 
                  Phys. Lett. B{\bf 357} (1995) 573.
\bibitem{CKN}      A.G. Cohen, D.B. Kaplan and A.E. Nelson,
                   Phys. Lett. B{\bf 388} (1996) 588.
\bibitem{ADM}     N. Arkani-Hamed, M. Dine and S.P. Martin,
                  Phys. Lett. B{\bf 431} (1998) 329.
\bibitem{BD}      P. Bin\'etruy and E. Dudas,
                  Phys. Lett. B{\bf 389} (1996) 503.
\bibitem{DP}      G. Dvali and A. Pomarol
                  Phys. Rev. Lett. {\bf 77} (1996) 3728;\\
                  R.N. Mohapatra and A. Riotto,
                  Phys. Rev. D{\bf 55} (1997) 1138;
                  Phys. Rev. D{\bf 55} (1997) 4262.
\bibitem{kurosawa} K. Kurosawa and N. Maekawa, Prog.Theor.Phys.
                  {\bf  102} (1999) 121.
\bibitem{LFV}     F. Borzumati and A. Masiero, Phys. Rev. Lett. 
                  {\bf 57} (1986) 961;\\
                  R. Barbieri and L. Hall, Phys. Lett. B{\bf 338}
                  (1994) 212;\\
                  J. Hisano, T. Moroi, K. Tobe, M. Yamaguchi and T. Yanagida,
                  Phys. Lett. B{\bf 357} (1995) 579; \\
                  J. Hisano, T. Moroi, K. Tobe and M. Yamaguchi,
                  Phys. Rev. D{\bf 53} (1996) 2442;\\
                  J. Sato and K. Tobe, hep-ph/0012333.
\bibitem{BKM}      M. Bando, T. Kugo and N. Maekawa, in preparation.
\end{thebibliography}
\end{document}